\documentclass[useAMS,usenatbib,fleqn]{mn2e}
\usepackage{graphicx, amssymb}
\usepackage[fleqn]{amsmath}
\usepackage{epsfig}
\usepackage{color}
\usepackage{times}
\usepackage{float}
\usepackage{mathrsfs}

\title{Toward {\it ab initio} extremely metal poor stars }

\author[J. Ritter et al.]
{Jeremy S.~Ritter$^1$, Chalence Safranek-Shrader$^{2,3}$, Milo\v s~Milosavljevi\'c$^1$,  and Volker~Bromm$^1$\\
$^1$Department of Astronomy, The University of Texas at Austin, Austin, TX 78712, USA\\
$^2$Department of Astronomy and Astrophysics, University of California at Santa Cruz, Santa Cruz, CA 95064, USA\\
$^3$Department of Astronomy, University of California at Berkeley, Berkeley, CA 94720, USA
}

\newcommand{\msun}{\mathrm{M_\odot}}
\newcommand{\Zsun}{\mathrm{Z_\odot}}
\newcommand{\pc}{\mathrm{pc}}
\newcommand{\kpc}{\mathrm{kpc}}

\newcommand{\Myr}{\mathrm{Myr}}
\newcommand{\kyr}{\mathrm{kyr}}
\newcommand{\cmcubed}{\mathrm{cm^{-3}}}
\newcommand{\erg}{\mathrm{erg}}

\newcommand{\Htwo}{\mathrm{H_2}}
\newcommand{\hii}{HII }
\newcommand{\SN}{\mathrm{SN}}
\newcommand{\cloud}{\mathrm{cloud}}

\newcommand{\KellerStar}{SMSS J031300.36-670839.3}
\newcommand{\CaffauStar}{SDSS J102915+172927}

\newcommand{\beq}{\begin{equation}}
\newcommand{\eeq}{\end{equation}}
\newcommand{\bea}{\begin{eqnarray}}
\newcommand{\eea}{\end{eqnarray}}

\begin{document}

\label{firstpage}

\maketitle
\topmargin-1cm

\begin{abstract}
Extremely metal poor stars have been the focus of much recent attention owing to the expectation that their chemical abundances can shed light on the metal and dust yields of the earliest supernovae.  We present our most realistic simulation to date of the astrophysical pathway to the first metal enriched stars.  We simulate the radiative and supernova hydrodynamic feedback of a $60\,\msun$ Population III star starting from cosmological initial conditions realizing Gaussian density fluctuations.  We follow the gravitational hydrodynamics of the supernova remnant at high spatial resolution through its freely-expanding, adiabatic, and radiative phases, until gas, now metal-enriched, has resumed runaway gravitational collapse.  Our findings are surprising: while the Population III progenitor exploded with a low energy of $10^{51}\,\text{erg}$ and injected an ample metal mass of $6\,M_\odot$, the first cloud to collapse after the supernova explosion is a dense surviving primordial cloud on which the  supernova blast wave deposited metals only superficially, in a thin, unresolved layer.  The first metal-enriched stars can form at a very low metallicity, of only $2-5\times10^{-4}\,Z_\odot$, and can inherit the parent cloud's highly elliptical, radially extended orbit in the dark matter gravitational potential.
\end{abstract}


\section{Introduction}
\label{sec:intro}

Understanding how the first generations of stars ended the cosmic dark ages is a key frontier in modern astrophysics and cosmology \citep[for reviews, see][]{Fan06,Furlanetto06,Morales10,Bromm11,Pritchard12,Bromm13}. Of particular interest are the initial steps in cosmic chemical evolution starting from an essentially pure hydrogen and helium Universe \citep{Karlsson13}. How did the first stars, the so-called Population~III (Pop~III), enrich the early Universe with the first heavy elements? 
Elucidating early star formation and metal enrichment is crucial for connecting the fossil record preserved in the Galactic metal-poor stars and the ancient dwarf galaxies in the Local Group to their formation sites at high redshifts \citep[for review, see][and references therein]{Frebel15}.

There is an ongoing debate concerning the physics responsible for the transition from the top-heavy initial mass function (IMF) predicted for metal-free Pop III stars \citep[see][and references therein]{Stacy16} to the low-mass-dominated IMF observed in metal-poor Pop II and chemically mature Pop I stars \citep{Chabrier03}.  What are the respective roles of metal fine structure line and dust grain cooling channels? The metal line cooling dominates thermal evolution at intermediate densities, too low to facilitate fragmentation of gas into low mass stars.  Dust cooling can allow fragmentation into low mass stars, but only if grains are abundant at these low metallicities---a theoretically uncertain proposition.  The debate centers on the specific roles of metal line and dust cooling \citep[for review, see, e.g.,][]{Milosavljevic16}. Does metal line cooling affect the final, high-density outcome of the fragmentation process?  If it does, we expect to detect its influence at a minimum `critical metallicity', which for fine structure carbon and oxygen line cooling is $Z_\mathrm{crit,fs} = 10^{-3.5}\,\Zsun$ \citep{Bromm01,Bromm03,Santoro06,Smith09}.  In comparison, the critical metallicity above which dust can explain the formation of the known metal-poor stars is $Z_\mathrm{crit,dust} \sim 10^{-6}\,\Zsun$, assuming a reference fraction of metal mass depleted in dust of $f_\text{dust}=0.22-0.24$ \citep[][see, also, \citealt{Omukai10,Ji14}]{Schneider06,Schneider12}.\footnote{\citet{Omukai10} reports a higher critical value $Z_\mathrm{crit,dust} \sim 10^{-5}\,\Zsun$ for $f_\mathrm{dust}\sim 0.01$ metal-to-dust depletion fraction.}
Observationally, there is a distinct lack of stars at metallicities $Z \lesssim 10^{-5}\,\Zsun$, but a handful of stars, such as \CaffauStar, with $Z = 10^{-4.35}\,\Zsun$ \citep{Caffau11}\footnote{For a fiducial solar metallicity of $Z_\odot=0.0153$ \citep{Caffau11}.}, have now been found that lie well below $Z_\mathrm{crit,fs}$.

The character of the very first metal and dust pollution is of course contingent on the physics of the initially metal-free Pop III stars and their explosions.  The principal model parameter is the Pop III stellar mass. Much effort is being put into attempts to constrain the statistics of this parameter with first-principles hydrodynamical simulations.  Recent high-resolution  simulations suggest that metal-free clumps in unpolluted cosmic minihalos fragment down to the protostellar mass scales of $\sim 1-10\,M_\odot$ 
\citep{Stacy10,Clark11,Greif11,Greif12}.  Accretion onto such protostars is eventually limited by radiative feedback \citep{Hosokawa11,Stacy12,Stacy16}.  The resulting stars should have final masses on the order of a few tens of solar masses. These massive stars rapidly convert hydrogen and helium into heavier elements in just a few million years \citep{Schaerer02} and explode as core-collapse supernovae \citep{Heger02}. 

The precise nucleosynthetic yields in the ejecta of core-collapse supernovae are not known from first principles, primarily because a theoretically uncertain  fraction of the supernova yield ends up trapped in the compact remnant \citep{Heger10}.  Three-dimensional simulations of the long-term evolution of core-collapse explosions indicate that the ejecta may be distributed anisotropically. The heavier elements (e.g., the iron group) that fall back onto the remnant in spherically-symmetric models are instead ejected in jets \citep{Wongwathanarat14}.  Intrinsic explosion anisotropy may imply some irreducible scatter in the heavier element abundances in enriched gas, even for similar mass supernova progenitors \citep{Sluder16}.  

These nonlinear effects can be so sensitive to the precise mass (and initial rotation) of the Pop III star that, in view of the theoretically uncertain Pop III IMF and rotational evolution \citep{Stacy11}, it is currently not possible to reliably predict the metal and dust yields of Pop III stars or the corresponding explosion kinetic energies.  Instead, attempts are being made to constrain the physics of Pop III star formation and evolution, and their metal enrichment, by surveying for chemically extreme low-mass stars, and then analyzing the abundance patterns in the most extreme ones that could have been enriched by \emph{single} Pop III supernovae.  For example, the absence of iron in the extremely metal poor \citet{Keller14} star has been interpreted as evidence for pollution by a single, low-energy (kinetic energy $\sim 10^{51}\,\text{erg}$), black-hole-producing Pop III supernova, with a stellar progenitor mass of $\sim 60\,M_\odot$ \citep[see, e.g.,][]{Ishigaki14,Kobayashi14,Takahashi14,Chen16}.  In contrast, an extremely metal poor star discovered in the Galactic bulge has been claimed to better fit a more energetic $\sim 10^{52}\,\text{erg}$ (`hypernova') explosion \citep{Howes15}.  Since dust is required for at least the final stage of gravitational fragmentation into low mass stars, there is a renewed push to study dust formation in Pop III supernovae \citep[e.g.,][]{Cherchneff10,Chiaki14,Chiaki15,Marassi14,Marassi15,Sarangi15}.

To date there has not been a direct numerical hydrodynamical demonstration of the single-supernova-to-low-mass-star pathway.  Can extremely metal-poor, low-mass stars really form in the aftermath of single Pop III supernovae?  It is easy to see how the Universe may fail to realize this pathway: the recovery time for re-condensation of gas in the wake of supernovae may be so long that the typical metal-enriched stars are always polluted by multiple Pop III explosions occuring in neighboring halos that ultimately merge with each other \citep[for inter-halo enrichment, see][]{Smith15}.  The enrichment by multiple Pop III supernovae could also be the predominant scenario if metals are so efficiently ejected from the host halos that a single halo must proceed through a sequence of Pop III-type explosions.  At the other end of the range of possibilities, the typical single Pop III could disperse its metals so compactly, with minimum dilution, as to immediately produce stars much more metal rich than the observed extremely metal poor stars.  Thanks to its complexity, the recycling of Pop III ejecta into the first metal-enriched stars remains an active research area \citep[e.g.,][]{Whalen08,Ritter12,Ritter15,Cooke14b,Jeon14,Sluder16}. 

Here, we attempt such a demonstration of the path toward star formation enriched by a single, low-energy Pop III supernova.  We present an {\it ab initio} cosmological hydrodynamical simulation involving a single Pop III supernova with a low explosion energy of $10^{51}\,\text{erg}$, proceeding until a metal-enriched pre-stellar core has formed in the cosmological box and the conditions imply imminent fragmentation into low-mass stars.  The simulation is a test of the single-supernova enrichment hypothesis for stars like that of \citet{Keller14}.  It also helps us understand precisely \emph{how} a smattering of Pop III ejecta gets combined with a much larger mass of metal-free gas, something that must happen if the system is to produce an extremely metal poor pre-stellar clump.  The simulation enables us to ask, on what initial orbits in the host dark matter halo do the first metal-enriched stars form?  The orbital extent could be related to the half-light radii of the ultra faint dwarf (UFD) galaxies.  The simulation substantially improves upon the idealized precursor simulations we reported in \citet{Ritter12} and \citet{Sluder16}.  It is the logical next step toward providing an end-to-end proof-of-concept for extremely metal poor star formation in the first metal-enriched cosmic objects.

The paper is organized as follows.  In Section \ref{sec:methods} we describe our numerical methodology and emphasize the novel elements introduced in this work.  In Section \ref{sec:results} we provide a descriptive account of the progression from a Pop III supernova to a metal-enriched pre-stellar clump.  We also offer our interpretation of how the clump acquired its extremely low metallicity.  In Section \ref{sec:discussion} we analyze our results and discuss the implications for interpreting extremely metal poor star discoveries and for explaining the kinematic structure of UFDs. Finally, in Section \ref{sec:summary} we summarize our main conclusions.

\section{Numerical methodology}
\label{sec:methods}

The initial setup of the simulation was the same as in \citet{Ritter12,Ritter15}.  We ran the adaptive mesh refinement (AMR) hydrodynamics code \textsc{flash} \citep{Fryxell00} with the multigrid Poisson solver of \citet{Ricker08} and 
dark matter particle smoothing and time-integration as in \citet{SafranekShrader12} and \citet{Ritter12}.  The $1\,\text{Mpc}$ comoving cosmological box was initialized at redshift $z = 145$ with nested Gaussian density fluctuations generated with the package \textsc{grafic2} \citep{Bertschinger01} using the {\it Wilkinson Microwave Anisotropy Probe} 7-year cosmological parameters   \citep{Komatsu11}.  The dark matter particle mass in the highest refined patch, where the gas would first collapse to high densities, was $M_\mathrm{DM} = 230\,\msun$, and the effective grid resolution (relative to the whole box at initialization) was $512^3$.

During the first, metal-free phase of the simulation, we tracked the formation of the principal coolant, the $\Htwo$ molecule, by integrating the standard ionic and molecular non-equilibrium chemical network \citep[see, e.g.,][]{SafranekShrader10}. At low densities and temperatures, we computed the cooling rate following the recommendation of \citet{Glover08}.  The metal-free phase lasted until a dark matter halo with a mass $\sim 10^6\,\msun$  and radius $\sim 175\,\pc$ virialized at redshift $z=19.5$ and the $\Htwo$-cooled, metal-free gas in the halo had collapsed to density $n_\mathrm{H}>10^3\,\text{cm}^{-3}$.  We interpreted this as heralding the imminent formation of the very first, metal-free star.  We replaced $60\,\msun$ of gas around the density maximum with an `active' (gravitating) Pop III star particle. This replacement capped the gas density around the maximum, where the star particle was inserted, at a ceiling computed consistent with mass conservation.

We ray-traced ionizing radiation from the star particle to compute the dynamical expansion of the \hii region over the star's lifetime of $3.5\,\text{Myr}$. The ionization state and temperature along each separate ray in the \hii region interior were evaluated by interpolating from values pre-computed with the code \textsc{cloudy} \citep{Ferland13} for a stellar surface blackbody spectrum with an effective temperature of $T_{\mathrm{eff}} = 10^{4.94}\,\mathrm{K}$  \citep{Schaerer02}. We tabulated the pre-computed data as a function of the photoionization parameter $\xi = F/n$, where $F$ is the local, optical-depth-attenuated flux \citep[see][]{Ritter12,Ritter15}.  For the ray-tracing calculation, we carried out finite volume mapping of the Cartesian AMR mesh onto a spherical mesh defined by the \textsc{healpix} scheme \citep{Gorski05}  partitioning the angular space into 3000 equal solid angle conical pixels and 100 logarithmically spaced radial bins. 
The stellar source emitted $Q = 4.795 \times 10^{49}$ ionizing photons per second, the expected stellar lifetime-average ionizing photon emission rate for a $60\,\msun$ star \citep{Schaerer02}.  The local ionizing flux in bin $b$ at distance $r$ from the source was $F_b =  (Q - \sum_{i \leq b}{V_i \alpha(T_i) n_i^2})/{4 \pi r^2},$ where $\alpha(T_i)$ is the temperature-dependent recombination rate and the sum is over all bins $i$ between bin $b$ and the ionizing source. For each grid cell $c$ that maps to conical radial bin $b$ we calculated the photo-ionization parameter $\xi_c = F_b / n_c$ and then linearly interpolated the tabulated solutions to update the temperature and ionization state in the cell. 

\begin{figure*}
\begin{center}
\includegraphics[width=0.32\textwidth]{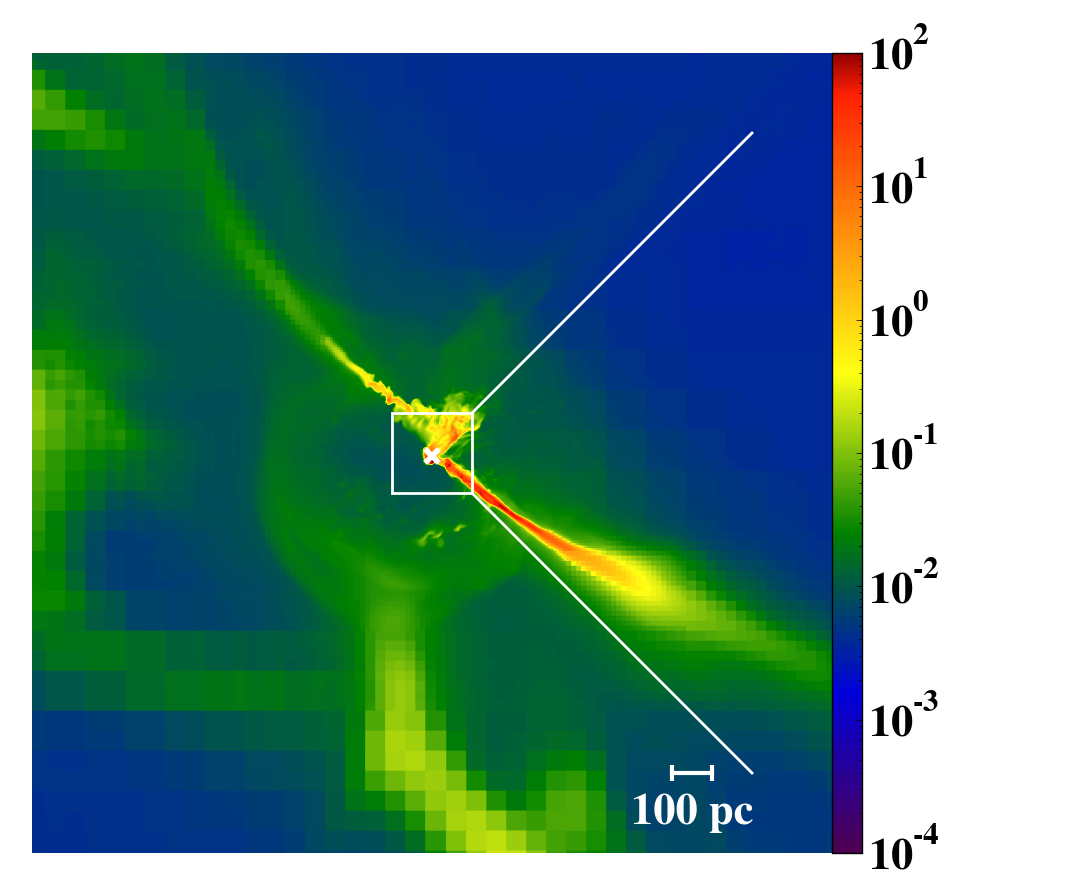}
\includegraphics[width=0.32\textwidth]{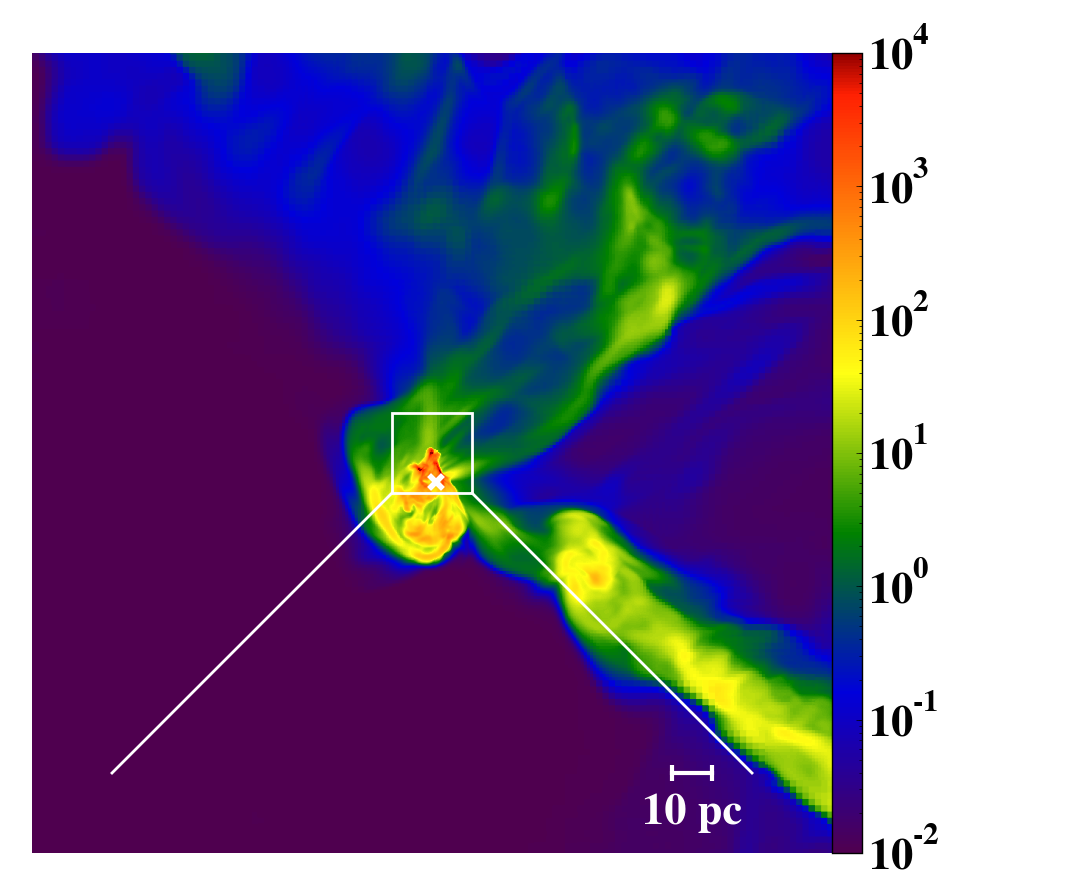} \\
\includegraphics[width=0.32\textwidth]{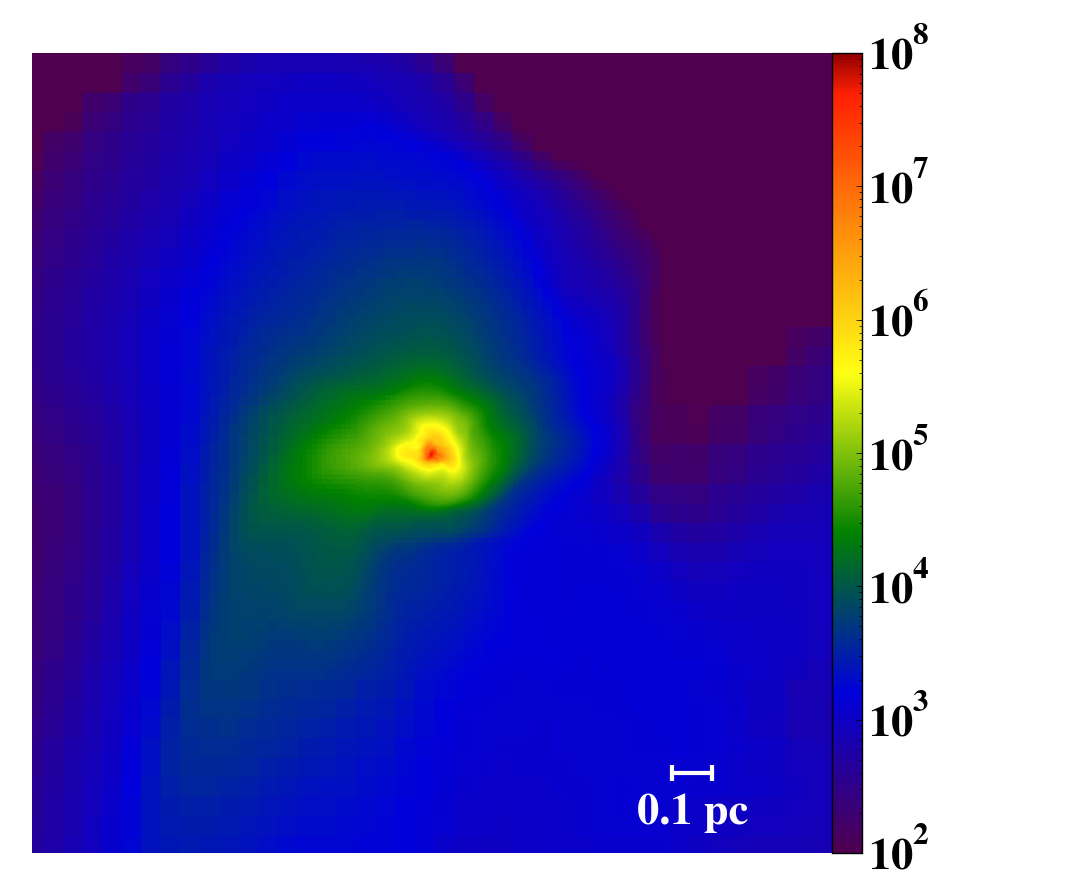}
\includegraphics[width=0.32\textwidth]{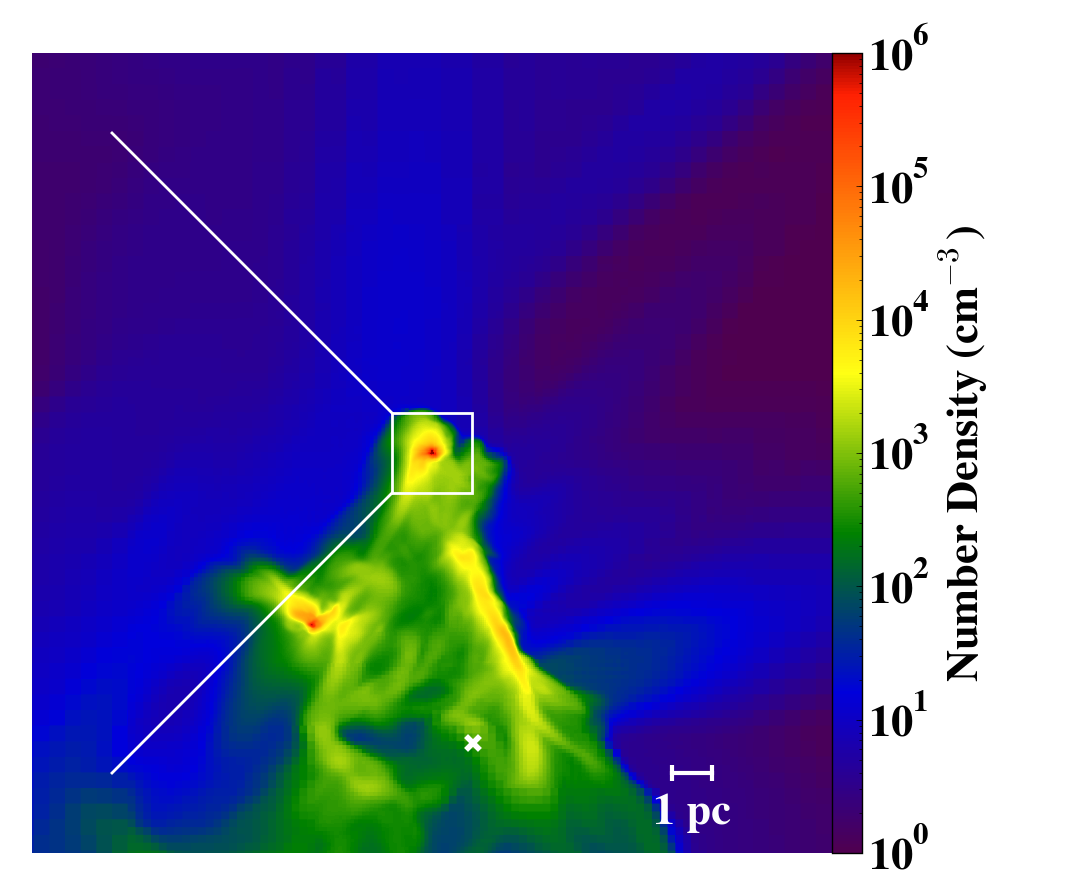}
\caption{Number density projections (density-weighted, color scale showing $m_\text{H}^{-1} \int\rho^2 dz/\int \rho dz$) centered on the densest gas cell at the end of the simulation, $13.7\,\Myr$ after the explosion. 
The top-left panel shows the cosmic web filaments feeding into the halo center.  The supernova ejecta bubble extends beyond the virial radius $r_\mathrm{vir}=175\,\pc$. The top-right panel shows that the dense neutral clouds that have survived photoionization are streaming into the halo center marked with a white ``$\times$''; the tip of a cosmic web filament is still a few dozen parsecs away.  The bottom-right shows dense, self-gravitating clumps at the apocenter of an elliptical gas flow pattern. 
Each clump contains a few hundred solar masses of gas.  The clumps loiter at density $n_\mathrm{H} \sim 10^4\,\cmcubed$ for several million years and then enter runaway collapse. The bottom-left panel shows the first runaway clump in which the collapse facilitated by metal line cooling rapidly increased the density above $n_\mathrm{H} > 10^8\,\cmcubed$.
\label{fig:morphology}}
\end{center}
\end{figure*}

\begin{figure*}
\begin{center}
\includegraphics[width=0.32\textwidth]{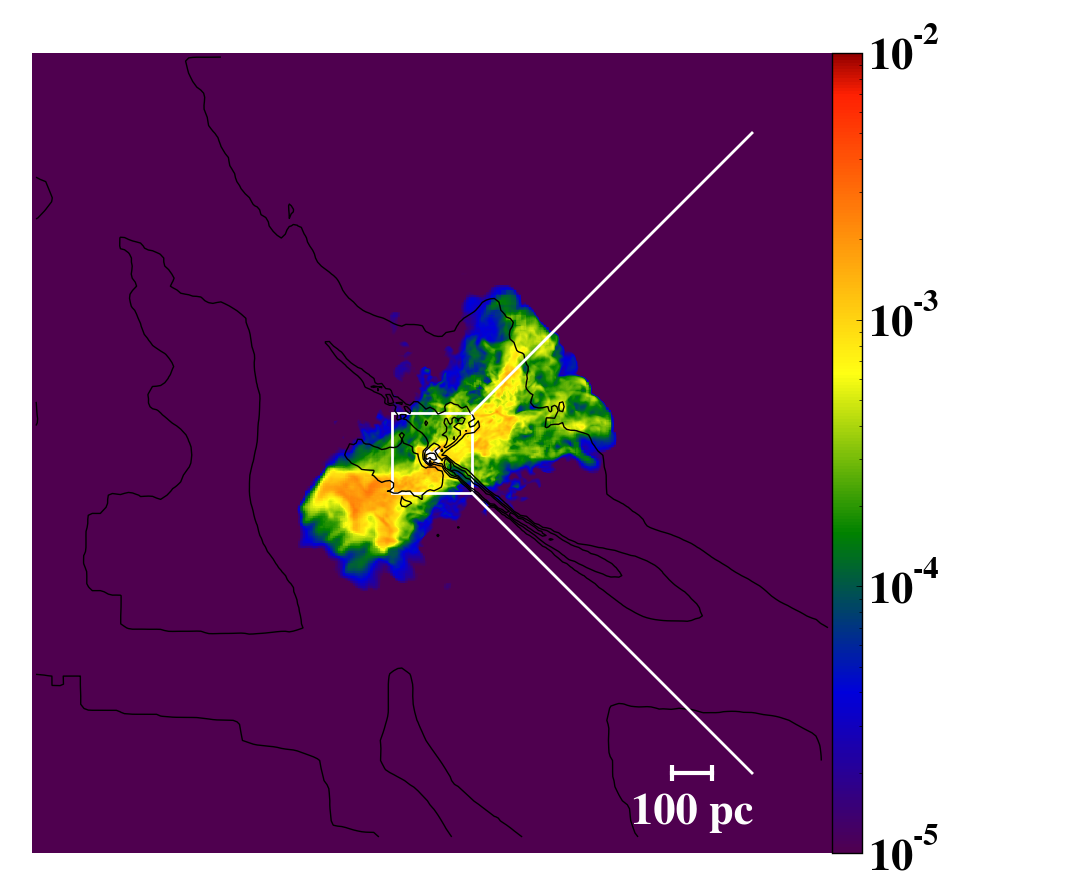}
\includegraphics[width=0.32\textwidth]{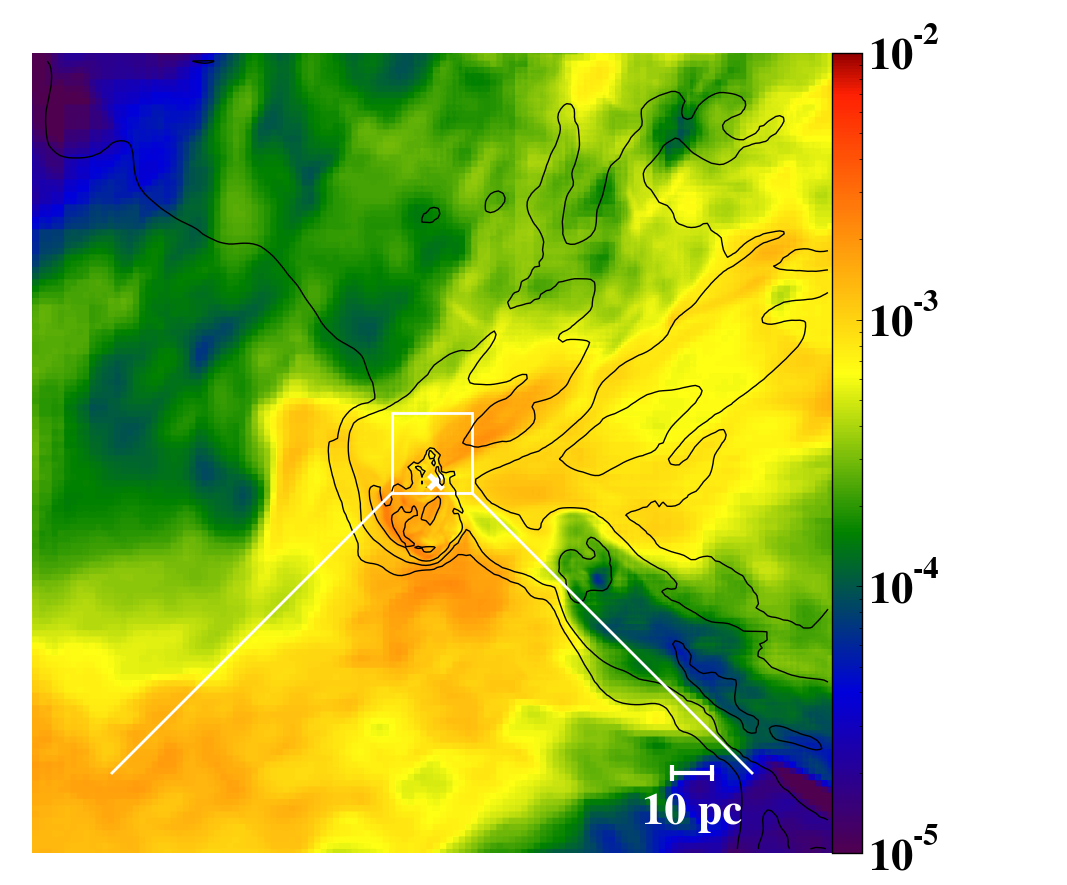} \\
\includegraphics[width=0.32\textwidth]{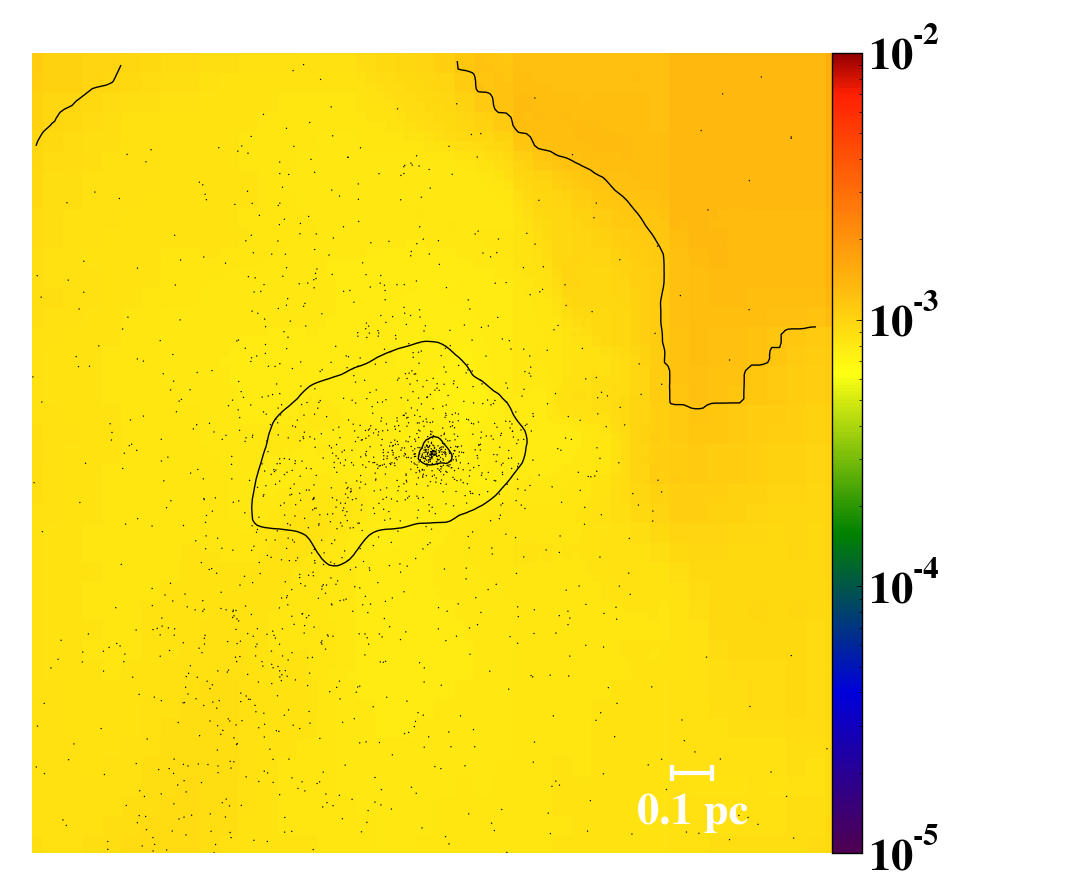}
\includegraphics[width=0.32\textwidth]{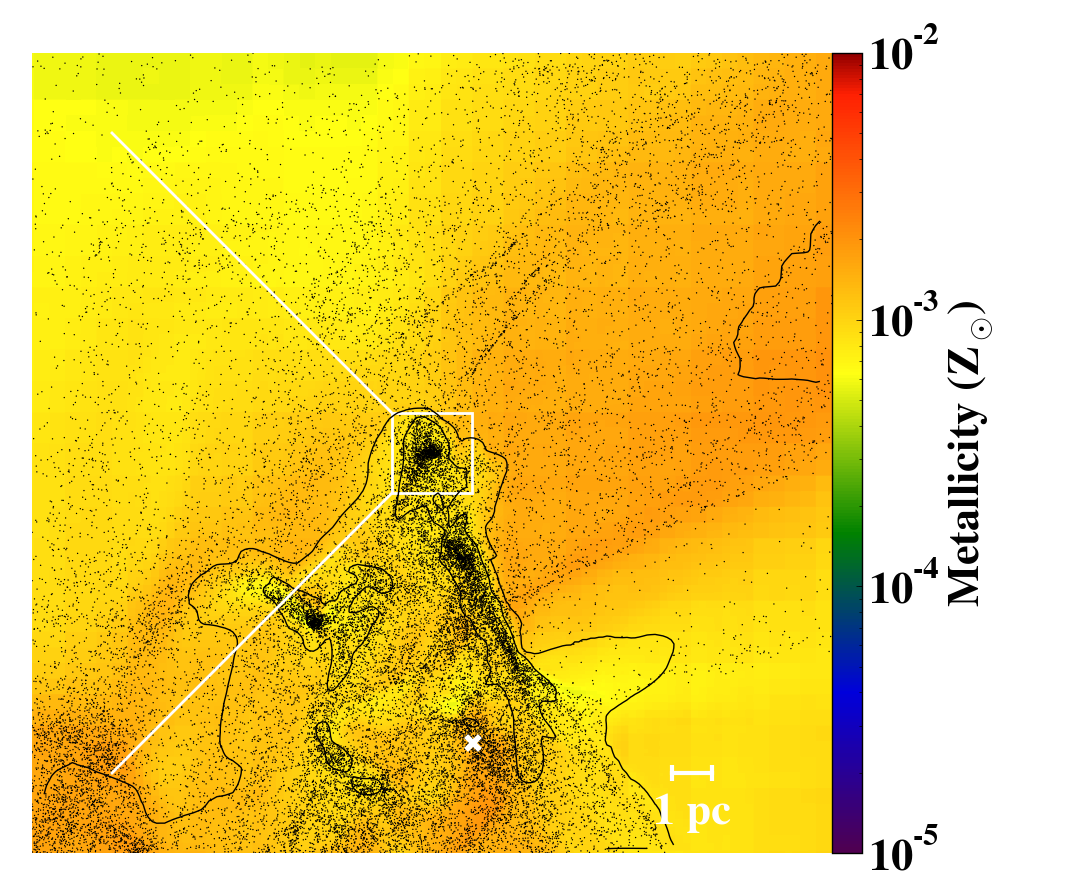}
\caption{The same as Figure \ref{fig:morphology}, but now showing metallicity projection (density-weighted, color showing $\Zsun^{-1} \int Z_\mathrm{scal}\rho dz/\int \rho dz$).  Here we overplot density contours to provide context for the clumps and filament locations. In the bottom panels we also overplot metal tracer particles.
\label{fig:metal_morphology}}
\end{center}
\end{figure*}

\begin{figure*}
\begin{center}
\includegraphics[width=0.49\textwidth]{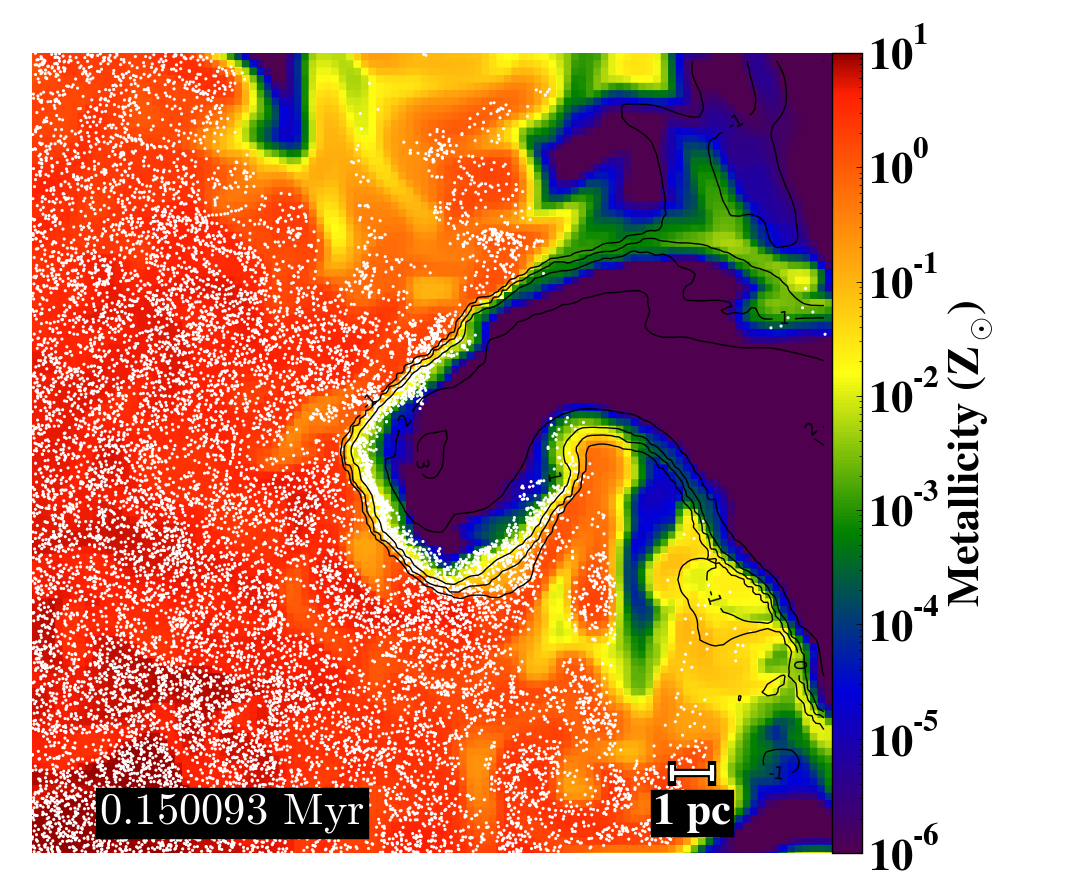}
\includegraphics[width=0.49\textwidth]{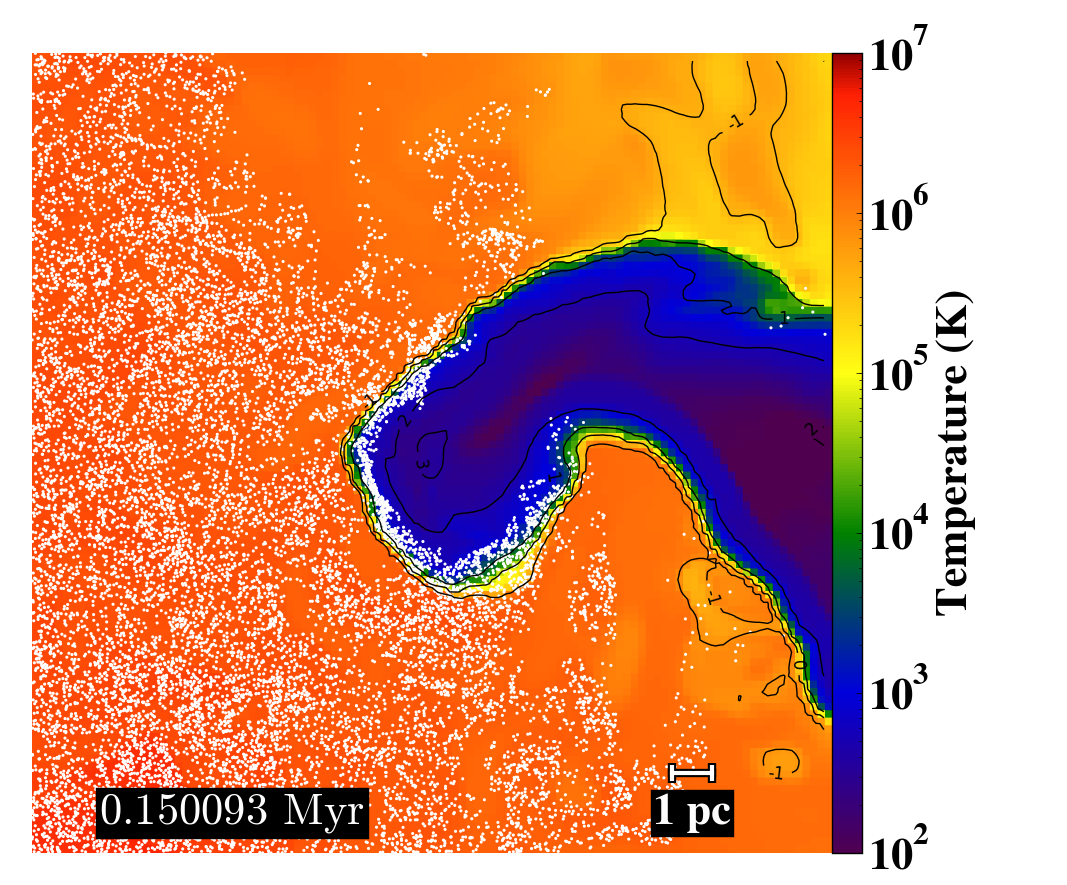}
\caption{A slice through the cold cloud that would ultimately collapse and fragment into metal-enriched pre-stellar clumps showing passive scalar metallicity (left panel) and temperature (right panel) with gas density contours overplotted.   Also overlplotted are Lagrangian passive ejecta tracer particles (white dots) in a $1\,\text{pc}$ thick slab containing the slice.  Note the superficial enrichment of the cloud as well as evidence of numerical passive scalar diffusion where the scalar metallicity is $>10^{-6}\,Z_\odot$ and yet no particles are present.
\label{fig:cloud_slice}}
\end{center}
\end{figure*}

When the star reached the end of its life we zoomed the mesh refinement to high resolution, from $\Delta x=0.7\,\text{pc}$ during the stellar lifetime to a minimum cell width of $\Delta x_\text{min}=0.02\,\text{pc}$, and inserted a supernova blast wave carrying $10^{51}\,\erg$ of kinetic energy.  The high spatial resolution was necessary to properly capture the reverse shock and avoid the notorious overcooling in under-resolved simulations \citep[see, e.g., the analysis in][]{Agertz13}.  Extrapolating from the model computations of \citet{Heger02}, we assumed that a $35\,\msun$ compact remnant was left behind and $25\,\msun$ of ejecta was launched outward in spherically symmetric fashion. For this simulation we assumed that metals in the ejecta accounted for $10\%$ (or $6\,\msun$) of the initial progenitor mass, which is $24\%$ of the ejecta mass ($Z_\mathrm{SN} = 12\,\Zsun$). The supernova remnant was inserted in the free-expansion phase by overwriting grid cells with a constant density, temperature, and metallicity within a radius of $R_{\SN} = 0.6\,\pc$ from the star particle. The velocity of ejecta was linear in radius $v(r) = v_{\mathrm{max}}\,r/ R_\SN$ and $v_{\mathrm{max}} = 2600\,\mathrm{km}\,\mathrm{s}^{-1}$. Ten million massless Lagrangian passive tracer particles were uniformly distributed in the ejecta to redundantly track their transport.  The advantage of Lagrangian tracers is that they are not subject to the same numerical diffusion in moving compositional discontinuities that affects the transport of the passively advected scalar metallicity variable.

After supernova insertion, we evaluated the cooling rate and ionization state of gas by interpolating from a new set of \textsc{cloudy} tables, now as a function of density, temperature, and metallicity, but with no radiation field. The tables were pre-computed for a purely atomic gas containing $\mathrm{C}$ and $\mathrm{O}$ in addition to the primordial species with relative abundance $[\mathrm{O}/\mathrm{C}] = -0.4$.  This value was chosen to be consistent with the abundances reported for carbon-enhanced metal-poor (CEMP) stars such as the star of \citet{Keller14}.\footnote{For the \citet{Keller14} star, \citet{Bessell15} report a measurement $[\mathrm{O}/\mathrm{C}] \approx 0$.  We were not aware of this more oxygen rich measurement when designing our simulation.}  Since $\Htwo$ cooling may still dominate at intermediate densities ($n_\mathrm{H} \sim 100\,\cmcubed$) and low metallicities ($Z \lesssim 10^{-2}\,\Zsun$) \citep{Glover14,SafranekShrader14}, we included additional cooling due to $\Htwo$ following \citet{Glover08} (assuming 3:1 ortho-to-para ratio).  To avoid the costly integration of the $\Htwo$ rate equation, in the final stage of the simulation we fixed the molecular abundance to $n_\Htwo  = 10^{-2}\,n_\text{H}$, a level consistent with detailed calculations of the Myr-time-scale $\Htwo$ abundance in recombining metal-poor flash-ionized regions \citep[e.g.,][]{Shapiro87,Kang92,Nakagura05,Johnson06,Yoshida07}. The thermodynamic update of specific internal energy $e$ was operator-split from the hydrodynamic update and was sub-cycled on sub-steps of a tenth of the equilibrium cooling timescale $t_{\mathrm{eq}} = \min(\rho e / \Lambda_{\Htwo} n_\mathrm{H} n_{\mathrm{H}_2},\rho e / \Lambda_Z n_\mathrm{H} n_Z)$.

We followed the blast wave for $13.7\,\text{Myr}$, at all times maintaining the volume affected by the blast wave at a maximum spatial resolution.  The adaptive refinement level was set such that the diameter of the blast wave was always resolved by at least $128$ grid cells and the entire region affected by the blast wave was fully contained within the highest-resolved AMR blocks.  During the final phase, when gaseous gravitational collapse resumed in the dark matter halo, we further refined the grid around the densest gas clumps, this time ensuring that the gas Jeans length was always resolved by at least $32$ cells.\footnote{The requirement that the thermal Jeans length be resolved by a minimum number of mesh elements is derived heuristically, not from {\it ab initio} numerical analysis.  The most careful analyses advise that the Jeans length should be resolved by at least $64$ cells, more than we require \citep[e.g.,][]{Latif13,Meece14}.}  To reduce the computational cost, during the final phase we degraded the mesh resolution in cells far from the halo center where the loss of resolution could have no impact on the dynamics in the self-gravitating clumps of interest.

\section{Results}
\label{sec:results}

Photoelectric heating from photoionization, which raised the gas temperature in the immediate vicinity of the Pop III star particle to $\sim 6\times10^4\,\mathrm{K}$, suddenly overpressured the ionized gas.  This drove a relatively dense shell ($n_\mathrm{H} \sim 100\,\cmcubed$) to distances $\sim 40 \text{--} 50\,\pc$.  The shell expanded isotropically except for in a few special directions in which there was the most pronounced inflow from the cosmic web.  There, embedded within the shell, remained denser ($n_\mathrm{H} \sim 10^3\,\cmcubed$) \emph{neutral} clumps---surviving fragments of the Pop III star's birth cloud. The failure to photoionize some of the densest metal-free clouds outside the ionizing sources' host halos was already seen in our previous published simulations \citep{Ritter12,Ritter15,Sluder16}; it had first been reported in the context of cosmic minihalos by \citet{Abel07}.  The failure to completely photoionize the densest clouds is also expected by scaling the most sophisticated analytical models of photoevaporating clouds \citep{Bertoldi90} to our metal-free, $\Htwo$-cooled clouds.  While the ionizing source was present, we did not compute $\Htwo$ cooling in the neutral clouds because molecules should have been wiped out by the strong local Lyman-Werner band $\Htwo$-dissociating flux.

The $3.5\,\text{Myr}$-old \hii region defined the hydrodynamic environment into which the supernova ejecta expanded upon insertion.  The reverse shock in the ejecta separated from the contact discontinuity at $\sim2\,\text{kyr}$ after insertion and reached the innermost ejecta after $\sim20\,\text{kyr}$.  The forward shock became radiative and formed a snowplow shell at $\sim150\,\text{kyr}$.  The blast wave partially ablated, but did not significantly disrupt the densest, persistently-neutral clouds at distances $\sim50\,\text{pc}$ from the explosion. The cloud fragments are visible in projection in the top-right panel of Figure \ref{fig:morphology}.   By the end of the simulation, the snowplow shell, also visible in the panel, has expanded to $\sim 0.5\,\text{kpc}$, over twice the virial radius of the halo. In the companion Figure \ref{fig:metal_morphology}, we 
show the corresponding distribution of metals (further discussed below).

Of particular interest is the interaction of the metal-bearing ejecta with the dense clouds, since, as we shall see, star formation can resume precisely when one such dense cloud falls toward the halo center.  At the end of the radiative phase the largest cloud had a central density $> 10^3\,\cmcubed$ and diameter $\sim 2\,\pc$ (see Figure \ref{fig:cloud_slice}).  By the time the supernova blast wave was passing the clouds, a fraction of the ejecta had already condensed in the blast wave's dense snowplow shell.  When the shell collided with the neutral clouds, the latter were temporarily shock-heated to $\approx 10^4\,\mathrm{K}$ but cooling by $\Htwo$ then quickly brought the temperature back down to $\sim 100\,\mathrm{K}$.  In this blast wave-cloud collision, a tiny fraction of the ejecta was deposited in a thin layer at the cloud boundary.  We detect this superficial deposition of metals on the surface of the metal-free cloud both in the passive scalar metallicity and the passive Lagrangian tracers.  Figure \ref{fig:cloud_slice} shows a slice through the cloud with the metallicity and temperature given in color scale and density contours as solid black lines.  The Lagrangian particles tracing ejecta in a thin slab containing the slice are shown.   Note the tracer particle concentration at the interface of the cold, dense cloud and hot, low density supernova remnant interior.

We are confident that the superficial metal deposition is a qualitatively correct prediction of the simulation, but some caveats are in order.  The boundary layer where the ejecta `sticking' to the cloud takes place is not adequately resolved: the forward shock in the cloud, the reverse shock in the blast wave, and the contact discontinuity that separates them are partially blended together.  This exacerbates the possibility of artificial passive scalar metallicity and internal energy diffusion across the blended discontinuity.  We cannot claim to resolve the correct physical hydrodynamics of the blast wave-ejecta interface.  For example, we do not resolve the Kelvin-Helmholtz instability that may develop there.  Consequently, we cannot be confident in the numerical convergence of the total metal mass that has been deposited on the metal-free cloud.  We describe how we attempt to exploit our dual ejecta tracking approach to constrain our prediction of the metallicity to a range of plausible values.

\begin{figure*}
\begin{center}
\includegraphics[width=0.49\textwidth]{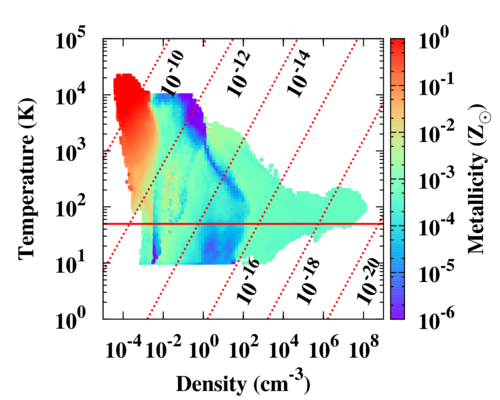}
\includegraphics[width=0.49\textwidth]{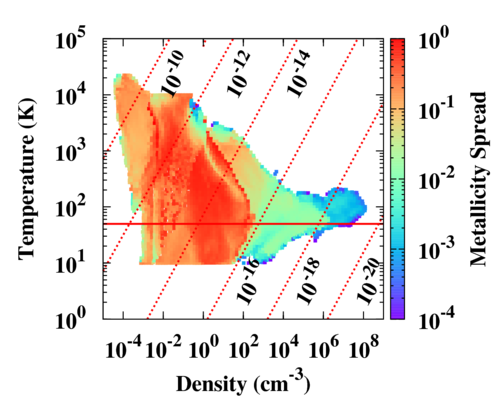}
\caption{Density-temperature phase diagrams for gas within $1\,\kpc$ of the minihalo at the end of the simulation, $13.7\,\Myr$ after the explosion. The color scale shows the mass-weighted average metallicity (left panel) and the relative log-metallicity dispersion (right panel). Dispersions near unity reflect unmixed, chemically heterogenous medium whereas lower values indicate homogenization. The densest gas with metallicity $\bar{Z}_\mathrm{scal} \approx 10^{-3.3}\,\Zsun$ has very low dispersion.  We also show the lines of constant entropy (dotted; see Equation \ref{eq:entropy} and Figure \ref{fig:metal_entropy}). The solid line is the cosmic microwave background temperature $T_\text{CMB}=53\,\mathrm{K}$ at redshift $z = 18.5$.
\label{fig:density_temperature}}
\end{center}
\end{figure*}

The metals deposited on the cloud surface did not initially mix with the cloud's metal-free interior because the interior was not turbulent.  Turbulent transport is a necessary (though perhaps not a sufficient) condition for compositional mixing to be possible \citep[see, e.g.,][and references therein]{PanLiubin12}. Our resolution was not sufficient to reproduce the turbulent cascade within the cloud.  Even if we had been able to simulate at the required resolution, we expect that such turbulence would have been very weak because there would not have been an agent to drive it after the blast wave has passed.  Cloud infall in the simulation can be described as follows.  The superficially-enriched cloud fell close to the halo center after $\sim 6 \text{--} 7\,\Myr$.  The infall tidally stretched the cloud into a narrow stream.  As the stream fluid flowed across the grid, artificial numerical diffusion in the propagating compositional discontinuity at the stream surface \citep[e.g.,][]{Ritter12} mixed the metals at the surface with the hydrogen and helium in the interior.  By the time the head of the stream made its closest approach to the halo center, the stream gas was homogenized at a low metallicity $Z_\mathrm{scal} \approx 10^{-3.3}\,\Zsun$, if measured from the passive scalar, and a factor of $3$ lower, $Z_\mathrm{part} \approx 10^{-3.8}\,\Zsun$ if measured from the passive Lagrangian tracers.  We expect that the true stream metallicity, if corrected for the varied numerical artifacts that likely affect the two values, is bracketed by the measured values.

The stream was on a non-radial orbit about the center of the minihalo dark matter potential.  After it made its closest passage by the halo center, the approximately isothermal tidal compression raised the density to $\sim 10^4\,\cmcubed$.  In a pattern familiar from the simulations of clouds and stars disrupted by massive black holes in galactic nuclei \citep[e.g.,][]{Burkert12,Guillochon13,Coughlin16}, after the pericenter passage the tidal stream is on an elliptical orbit and on a collision course with its own tail that has not yet reached the pericenter.  The diameter of the elliptical orbit (major axis) is $\approx 10\,\pc$.  Interestingly, runaway gravitational collapse in the tidally-compressed stream gas takes place as the gas motion stalls at the apocenter of the ellipse.  That a pair of dense clumps have broken from the head of the stream can be seen in the bottom-right panel of Figure \ref{fig:morphology}. The peak density in the clumps loitered at $\lesssim 10^5\,\cmcubed$ for  $\approx 6\,\Myr$.  After the loitering period, as the density increased and temperature dropped, atomic fine-structure line cooling enabled gravitational instability and runaway collapse of the densest clump to density $>10^8\,\cmcubed$. This can be seen in the bottom-left of Figure \ref{fig:morphology} recorded at redshift $z = 18.5$, only $13.7\,\Myr$ after the supernova explosion.

The top two panels of Figure \ref{fig:metal_morphology} show that the distribution of metals is not homogenous throughout the volume affected by the supernova blast wave.  The most metal-rich gas is the low-density reverse-shocked interior of the ejecta.   However the densest clumps in the bottom panels of Figure \ref{fig:metal_morphology}, which is where gravitational instability sets in, are quite homogenous, something that could in part be a numerical, finite-resolution artifact of the numerical diffusion of the passive scalar metallicity.

\section{Analysis and discussion}
\label{sec:discussion}

\subsection{Metal enrichment}
\label{sec:mixing}

In Figure \ref{fig:density_temperature} we show density-temperature phase diagrams for gas within a large $1\,\kpc$ sphere around the halo center. In the left panel we color the bins with the metallicity in solar units ($\Zsun$). The warm gas at temperature $\sim 10^4\,\mathrm{K}$ is the diffuse ($n_\mathrm{H} \sim 10^{-4} \text{--} 10^{-3}\,\cmcubed$) and very metal rich (super-solar) bubble left behind by the supernova. The unpolluted gas ($Z_\mathrm{scal} \leq 10^{-6}\,Z_\odot$) also at temperature $\sim 10^4\,\mathrm{K}$ and at density $n_\mathrm{H} \lesssim 1\,\cmcubed$ is the photoheated intergalactic medium (IGM). The narrow strip of low metallicity medium that extends to temperature $\sim 100\,\mathrm{K}$ and density $n_\mathrm{H} \sim 100\,\cmcubed$ is the cosmic web filament. At densities higher than $n_\mathrm{H} \sim 100\,\cmcubed$ we find the metal-enriched cloud and collapsed clump with uniform metallicity $Z_\mathrm{scal} \approx 10^{-3.3}\,\Zsun$. 

%
%

The right panel of Figure \ref{fig:density_temperature} is the same plot but now colored to show the metallicity spread.  We define the spread as the mass-weighted standard deviation of $\log Z_\mathrm{scal}$ in units of the mass-weighted average value of $\log Z_\mathrm{scal}$.  The low density gas ($n_\mathrm{H} < 100\,\cmcubed$) has a large metallicity scatter and is composed of unmixed pockets of metal-rich ejecta and pristine metal-free gas. The spread in the enriched cloud, with density $n_\mathrm{H} > 100\,\cmcubed$, is small in comparison.  The spread is almost negligible in the collapsed clump that has density $n_\mathrm{H} > 10^4\,\cmcubed$. The decrease in metallicity spread in the collapsed clump relative to the uncollapsed clumps is a consequence of physical, but also numerical effects. Regardless of the numerical artifacts, self-gravitating gaseous collapse \emph{is} expected to chemically homogenize the gas \citep[see, e.g.,][]{BlandHawthorn10,Feng14}. Taken at face value, the simulation leads us to predict that the stars that would form in the gas clump should have a common metallicity and equal fractional abundances, though, as we showed in  \citet[][]{Sluder16}, the abundances may not reflect the monolithic, explosion-averaged abundances of the supernova ejecta.

We present a simple model for how an initially metal-free cloud acquires a net metallicity through blasting by supernova ejecta. The cloud's cross section can be approximated with a disk of cross-sectional area $\pi r^2_\cloud$. If the cloud is at a distance $d_\cloud$ from the explosion, the surface area of the blast wave when it reaches the cloud is $4 \pi d^2_\cloud$. The mass of metals ejected is the total supernova fractional mass yield, $Z_\SN$, times the total ejecta mass, $M_\SN$. Some fraction $f_\mathrm{eff}$ of the metals ejected in the cloud's direction are deposited on the cloud's surface. The net metallicity of the cloud can be estimated as
\beq
Z_\cloud = f_\mathrm{eff} \frac{M_\SN Z_\SN}{M_\cloud} \frac{r^2_\cloud}{4 d^2_\cloud} .
\label{eq:cloud_metallicity}
\eeq
Relating the cloud's radius to its mass $M_\cloud$ and density $n_\cloud$, we get
\beq
Z_\cloud = f_\mathrm{eff} \frac{M_\SN Z_\SN}{4 M_\cloud d^2_\cloud} {\left( \frac{3 M_\cloud}{4 \pi m_\mathrm{H} n_\cloud}\right) }^{2/3} .
\eeq
For a cloud with mass $M_\cloud = 1000\,\msun$ and average density, $n_\cloud = 100\,\cmcubed$, we get an approximate cloud size $r_\cloud \approx 5\,\pc$. If we choose 10\% for the capture efficiency, $f_\mathrm{eff} = 0.1$, and substitute the distance to the cloud, $d_\cloud = 50\,\pc$, we obtain $Z_\cloud \approx 10^{-4}\,\Zsun$. This back-of-the-envelope calculation demonstrates that cloud surface enrichment by supernova blasting can be one of the enrichment mechanisms responsible for extremely metal poor star formation. 

\begin{figure}
\begin{center}
\includegraphics[width=0.49\textwidth]{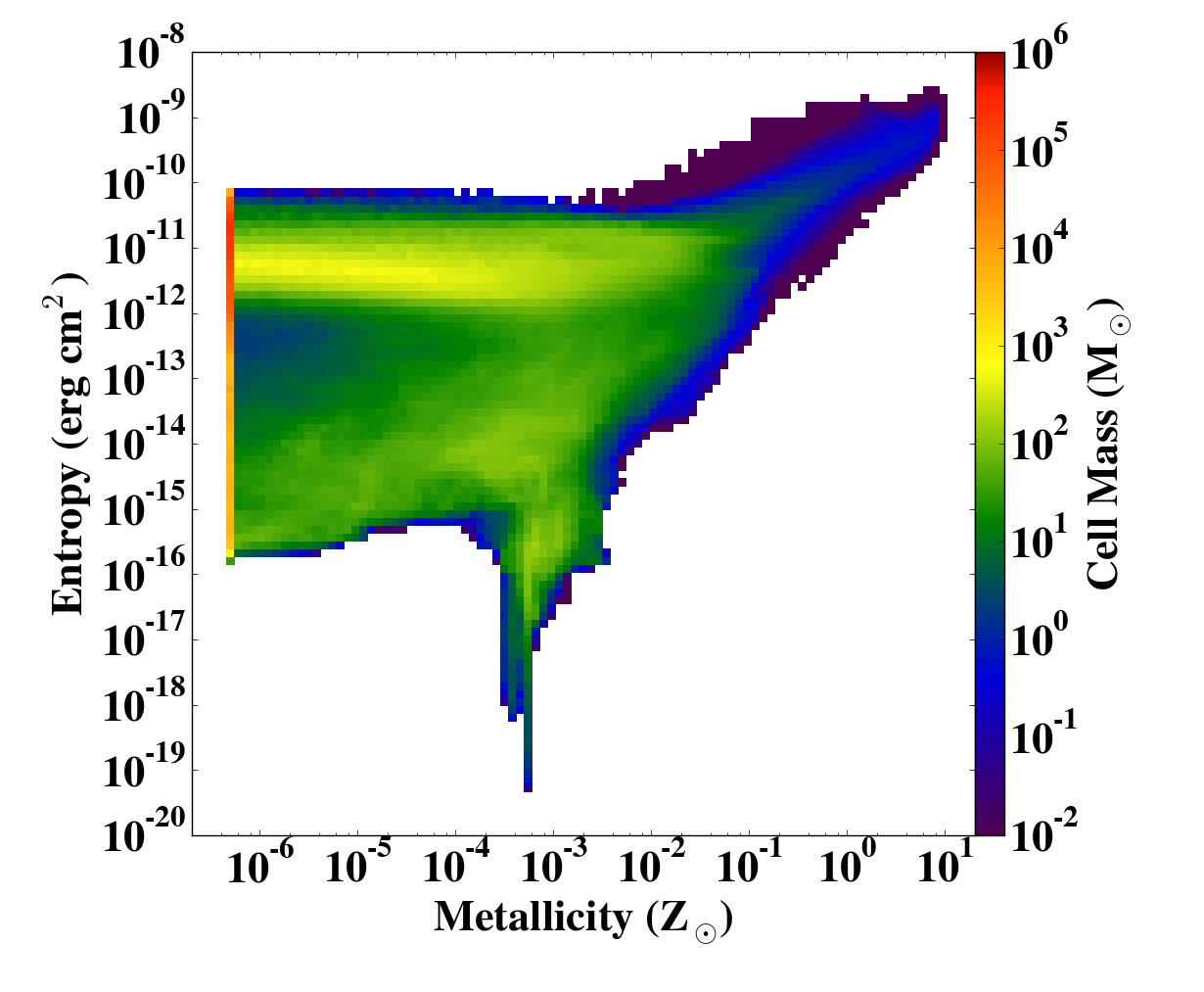}
\caption{Metallicity-entropy phase diagram for gas within $1\,\kpc$ of the minihalo at the end of the simulation, $13.7\,\Myr$ after the explosion. Entropy here is as defined in Equation \ref{eq:entropy}.  Very low metallicity gas $Z_\mathrm{scal} \leq 10^{-6}\,\Zsun$ was placed in the leftmost column. The color scale is the total gas mass contained in each two-parameter bin. The unmixed, diffuse, shocked ejecta, some of which escaped the halo, are on the upper-right. The shock-heated dense clouds that were blasted with metals have cooled and homogenized near metallicity $\sim10^{-3.3}\,\Zsun$ and then undergo gravitational collapse to form the low-entropy peninsula in the bottom-middle.
\label{fig:metal_entropy}}
\end{center}
\end{figure}

\subsection{Cooling thresholds}
\label{sec:cooling}

In Figure \ref{fig:metal_entropy} we plot metallicity and entropy for gas within the halo and nearby IGM. We parametrize the entropy using the argument, up to a constant factor, of the entropy logarithm,
\beq
S=\frac{k_\mathrm{B} T}{n^{2/3}}\ \ \ (\text{erg}\,\text{cm}^2) .
\label{eq:entropy}
\eeq
The high-entropy, metal-rich tail in the upper right of the plot is the unmixed, reverse-shocked supernova ejecta.  The lack of gas at high metallicities and low entropies on the lower right reflects geometric dilution in the dispersal of supernova ejecta. It may also have to do with a resolution-limited maximum density that metal-enriched gas gets compressed to in the thin snowplow shell.

The self-gravitating, collapsing clump appears as the low-entropy peninsula in the lower middle of the plot.  The peninsula is narrow around $Z_\mathrm{scal} \sim 10^{-3.3}\,Z_\odot$ (or 3 times lower, $Z_\mathrm{part}\sim10^{-3.8}\,Z_\odot$, if measuring metallicity with Lagrangian tracer particles) because the collapsing gas is chemically homogeneous.  Outside of the peninsula, the plot exhibits an entropy floor at $S\sim 10^{-16}\text{--}10^{-15}\,\text{erg}\,\text{cm}^{2}$.  Metal-free gas has been lumped into the lowest metallicity bin at $Z_\mathrm{scal} = 10^{-6}\,\Zsun$.   The low-entropy (i.e., cold and dense) metal-enriched gas clump loitered at the entropy floor for several million years before dropping further in entropy, i.e., entering runaway gravitational collapse facilitated by cooling.  It is not surprising that the entropy drop and the accompanying gravitational collapse happened in the most metal rich gas ($Z_\mathrm{scal}\sim 10^{-3}\,Z_\odot$) lying on the entropy floor, because it is in that gas that the metallic fine structure line cooling is the strongest and the net cooling time due to $\Htwo$ and metals is the shortest.

Consider a gravitationally-bound, metal-enriched gas clump with temperature $\sim 100\,\mathrm{K}$ collapsing isothermally at constant metallicity and $\Htwo$ abundance.  The $\Htwo$ as well as C and O fine structure line cooling 
times are $t_{\mathrm{cool},i} = \frac{3}{2} k_B T/\Lambda_i n_i$, where $i=(\Htwo,Z)$ and the free fall time is $t_\mathrm{ff} = (G \rho)^{-1/2}$. 
 Each cooling process has a characteristic critical level-thermalization density $n_\mathrm{crit}$ above which the volumetric cooling rate scales no longer as $\propto n_\mathrm{H}^2$, but as $\propto n_\mathrm{H}$, and the cooling time saturates toward higher densities.  At temperatures $\sim100\,\text{K}$, the critical density for $\Htwo$ is $n_{\mathrm{crit},\Htwo}\sim100\,\cmcubed$, whereas for C and O is substantially higher, $n_{\mathrm{crit},Z}\sim 10^6\,\cmcubed$.  Just above the $\Htwo$ critical density molecular cooling is inefficient, and the metal cooling, with its time scale on the order of $\sim 1$--$100\,\text{Myr}$, is still subdominant and comparable in magnitude to molecular cooling.   For the specific conditions in the simulation, the three time scales ($t_{\mathrm{cool},\Htwo}$, $t_{\mathrm{cool},Z}$, $t_\mathrm{ff}$) are roughly equal at density $n_\mathrm{H} \sim 10^4\,\cmcubed$. This explains why the densest gas in the simulation loitered at $\sim 10^4\,\cmcubed$ for $\approx 6\,\Myr$ before resuming collapse.

\begin{figure*}
\begin{center}
\includegraphics[width=0.33\textwidth]{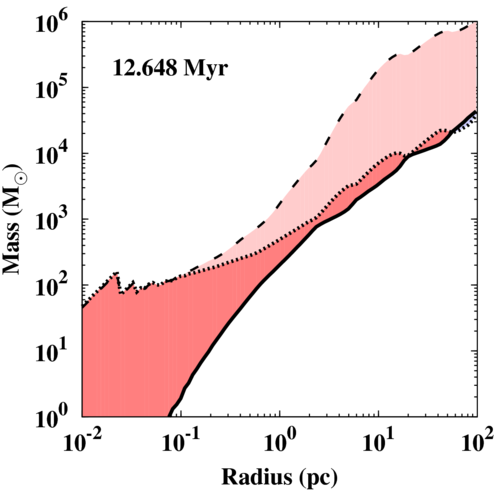}
\includegraphics[width=0.33\textwidth]{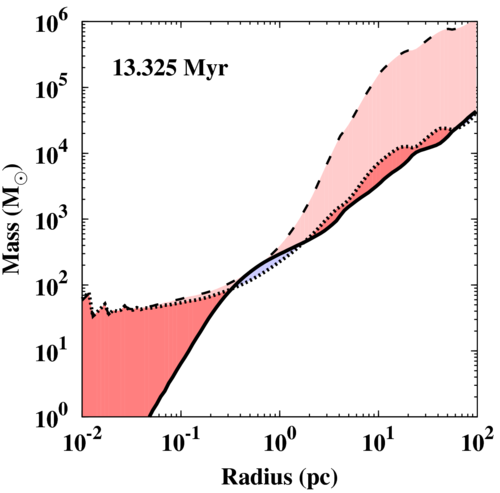}
\includegraphics[width=0.33\textwidth]{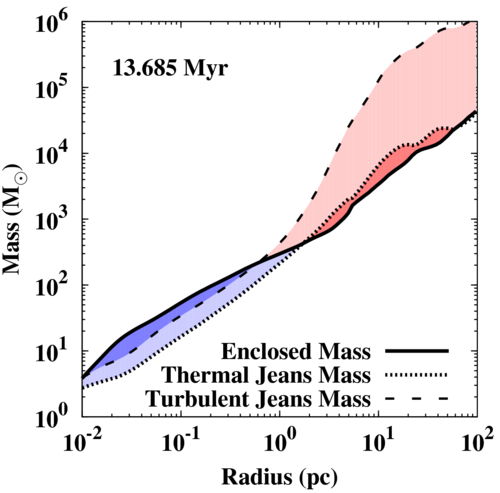}
\caption{Mass profile plots of the enclosed mass as well as the thermal (Eq.~\ref{eq:thermal_jeans}) and turbulent (Eq.~\ref{eq:turbulent_jeans}) Jeans masses, all calculated using averages within a sphere centered on the densest cell.
We shade the region red, light blue, or dark blue, depending on whether neither or only the thermal criterion, or in turn both criteria are satisfied.
 The left panel at $12.6\,\Myr$ is before the onset of collapse, when the gas density loitered at $n_\mathrm{H}\sim10^4\,\cmcubed$ and the enclosed mass was well below the Jeans mass (minimum cell size $\Delta x = 0.02\,\pc$). The middle panel at $13.3\,\Myr$ is on the cusp of runaway gravitational collapse when the enclosed mass has exceeded the thermal Jeans mass on radial scales of $\sim 1\,\pc$ ($\Delta x = 0.01\,\pc$). The panel on the right at $13.7\,\Myr$ is when the gas has entered runaway collapse and the enclosed mass has exceeded the thermal and turbulent Jeans masses on all scales shown between $0.01$ and $1\,\pc$ ($\Delta x = 4 \times 10^{-4}\,\pc$).
\label{fig:jeans}}
\end{center}
\end{figure*}

When the cooling time in a Jeans-unstable clump becomes shorter than the free fall time, gravitational contraction can occur on a dynamical time scale.   Shock compression in the gravitational infall may further shorten the cooling time and amplify short-wavelength density perturbations.  The perturbations may be necessary for gravitational fragmentation into low mass protostellar cores (at densities still higher than attained in the present simulation). We observe a clear loitering phase at $n_\mathrm{H} \sim 10^3$--$10^4\,\cmcubed$ where the drop of the $\Htwo$ cooling time under compression has saturated at a few million years and rises relative to the free fall time.  With increasing metallicity, the metallic fine structure line cooling, with its substantially higher critical density, picks up cooling closer to where the $\Htwo$ cooling saturates; this reduces the loitering duration compared to metal-free conditions.

\subsection{Gravitational instability}
\label{sec:instability}

Consider a cold, dense gas clump at the threshold of gravitational instability. The thermal Jeans mass is 
\beq
M_\mathrm{J} \sim \frac{c^3_\mathrm{s}}{\sqrt{G^3 \rho}} ,
\label{eq:thermal_jeans}
\eeq
where $c_\mathrm{s}$ is the isothermal sound speed.
The turbulent Jeans mass can be obtained by replacing the thermal sound speed with an effective velocity dispersion $v_\mathrm{eff}$ that combines in quadrature the isothermal sound speed and the turbulent velocity dispersion
\beq
M_\mathrm{J,turb} \sim \frac{v^3_\mathrm{eff}}{\sqrt{G^3 \rho}}.
\label{eq:turbulent_jeans}
\eeq
where
\beq
v^2_\mathrm{eff} = c^2_\mathrm{s} + v^2_\mathrm{turb}
\eeq
and $v_{\rm turb}$ is the mass-weighted standard deviation of the velocity component along any fixed axis.

Gas clumps collapsing under self-gravity should satisfy the turbulent Jeans criterion $M_\mathrm{enclosed} > M_\mathrm{J,turb}$. Figure \ref{fig:jeans} shows the spherical enclosed mass and the thermal and turbulent Jeans masses as a function of radius.  In the left panel which shows the gas clump during the loitering phase, before gravitational instability has set in, we notice that neither criterion is met. In the middle panel, $675\,\kyr$ later, the mass enclosed within radius $r \sim 1\,\pc$ has marginally exceeded the thermal Jeans mass. The gas does not immediately become gravitationally unstable after crossing this threshold, however, because of the additional turbulent velocity dispersion with Mach numbers just above unity.  In fact, we see in the right panel that it takes an additional $350\,\kyr$ before both Jeans criteria are satisfied and the collapse can proceed to smaller scales, down to the minimum cell size ($\Delta x = 4 \times 10^{-4}\,\pc$).

It is interesting that second generation stars will form in a kinematic state different from that of the first metal-free star.  Pop III stars form very close to dark matter potential minima because dark matter gravity drives gas compression down to very small scales $\lesssim1\,\text{pc}$ \citep[e.g.,][]{Bromm02}. Here, however, the metal-enriched pre-stellar clump poised to form Pop II stars can go into runaway gravitational collapse on an orbit that resides farther from the dark matter potential minimum (see the bottom-right panel of Figure \ref{fig:morphology}).  This is a consequence of the removal of gas from the halo center by \hii region heating and supernova ram pressure, as well as the rapid cooling in the clouds that fall toward the center. Since the cooling is rapid enough for the clouds to lose thermal energy before their kinetic energy can be dissipated, the new stars will be placed on large elliptical orbits, in this case $\sim10\,\text{pc}$, in agreement with the half-light radii of the very faintest UFDs \citep[e.g.,][]{Bechtol15,DrlicaWagner15,Koposov15}.

\subsection{The relation to extremely metal-poor stars}
\label{sec:populations}

The carbon-enhanced metal-poor stars, or CEMP stars in the nomenclature of \citet{Beers05}, are identified by having carbon-iron abundances [C/Fe]~$> 0.7$. They become increasingly common and more carbon-rich for iron abundances decreasing below [Fe/H]~$< -1$. The unenhanced, carbon-normal metal-poor stars have iron abundances that reach as low as [Fe/H]~$\sim -5$ \citep{Caffau11,Norris13}. The most extreme examples of carbon-rich stars, with [C/Fe]~$> 3$, are also the most extreme iron-poor stars with [Fe/H]~$< -5$ \citep{Christlieb02,Frebel05,Keller14}. Some carbon-rich iron-poor stars in the range $-3 <$~[Fe/H]~$< -1$ also have enhancements of the more rare s- and r-process neutron capture elements \citep{Sneden96, Cayrel01,Frebel07}, while below [Fe/H]~$< -3$, the carbon-rich stars show few signs of neutron-capture enhancement \citep{Aoki10}. The high degree of chemical variability in the most iron-poor stars tells us that the relative contributions from different classes of nucleosynthetic sources had not yet been homogenized when the stars formed. There is a growing class of chemically primitive stars with [Fe/H]~$< -5$ in which we clearly see the fossil imprint of individual nucleosynthetic sources, possibly even a single progenitor supernovae \citep{Christlieb02,Frebel05,Keller14}.  If the first stars predominantly exploded as core-collapse supernovae, each enriching gas with large amounts of $\alpha$-elements but relatively little iron, could they explain the universal carbon enhancement seen at [Fe/H]~$< -5$ \citep{Norris13}? 

\begin{figure}
\begin{center}
\includegraphics[width=0.45\textwidth]{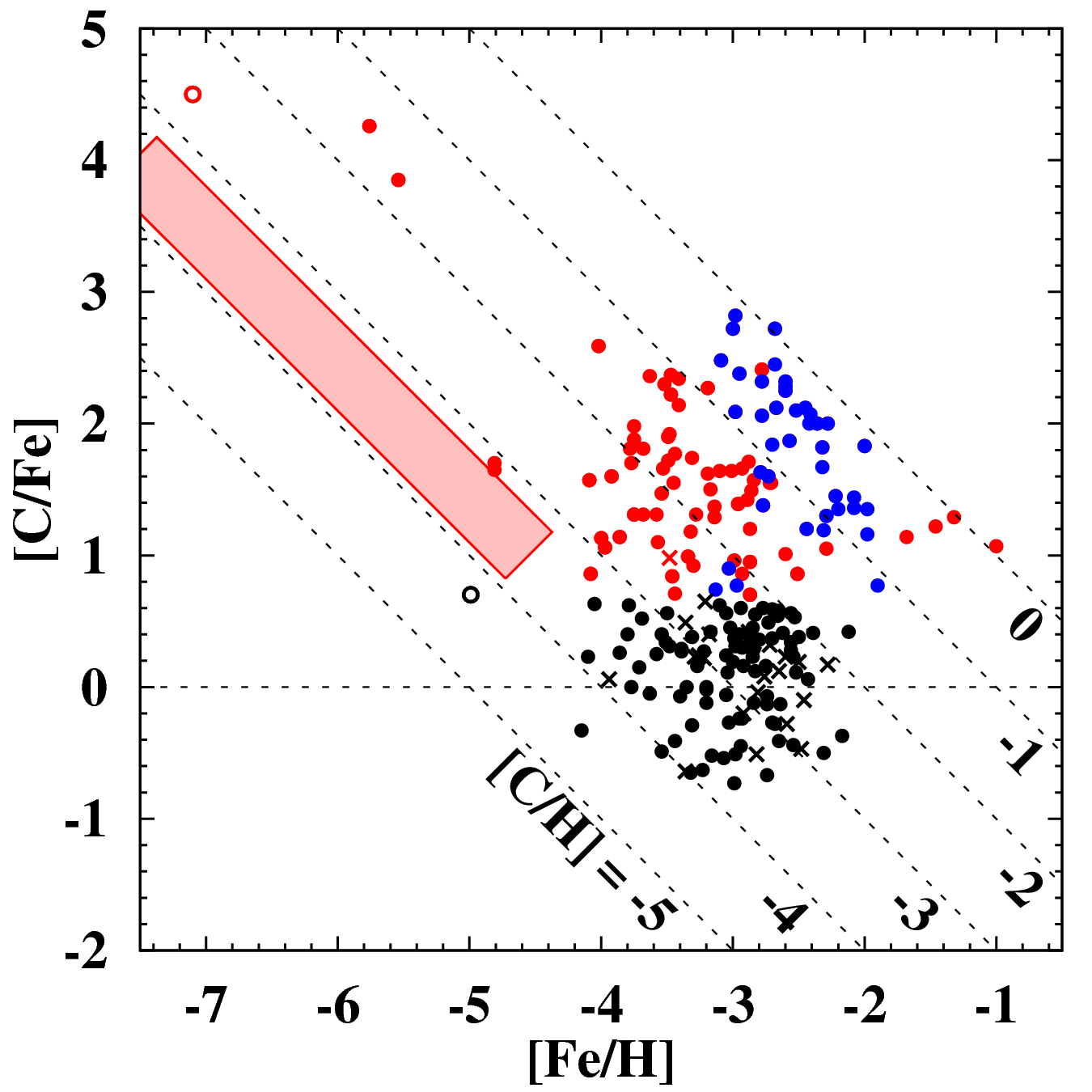}
\caption{Chemical abundance log-ratios [Fe/H] and [C/Fe] in the known metal-poor stars compiled from \citet{Caffau11}, \citet{Yong13}, \citet{Norris13}, \citet{Keller14}, and \citet{Howes15}.  The stars are colored depending on whether they are normal (black) or carbon-rich without (red) or with neutron capture enhancement (blue). Metal-poor stars from the Galactic bulge are identified using crosses, while all other stars are identified using circles. The carbon-rich, iron-poor star \KellerStar\ \citep{Keller14} is an open red circle. The carbon-normal, iron-poor star \CaffauStar\ \citep{Caffau11} is an open black circle. The diagonal dashed lines trace constant [C/H], which, assuming [C/O]~$\approx 0$, is approximately equal to the metallicity. The shaded region $-3.8<$~[C/H]~$<-3.3$ represents the stars that could have formed in our simulated gas clump assuming a range of carbon enhancements $1 <$~[C/Fe]~$< 4$. 
\label{fig:carbon_iron}}
\end{center}
\end{figure}

The stars which could form from the dense, metal-enriched gas clump in our simulation would have a metallicity in the range $(10^{-3.3}-10^{-3.8})\,\Zsun$, consistent with the canonical critical metallicity $Z_\mathrm{crit,fs} = 10^{-3.5}\,\Zsun$ for low-mass star formation in the absence of dust. In Figure \ref{fig:carbon_iron} we plot the carbon-to-iron ratio for known iron-poor stars with [Fe/H]~$< -2$ \citep{Yong13,Norris13}. The black points are carbon-normal stars with [C/Fe]~$< 0.7$. The red points are carbon-rich stars with [C/Fe]~$>0.7$, but those that show no sign of neutron-capture enrichment (i.e., CEMP-no). The blue points are carbon-rich stars that do show signs of additional s- and/or r-process enrichment, using the criteria of \citet{Beers05} for CEMP-r, CEMP-s, and CEMP-r,s stars. The most iron-poor star currently known\ \KellerStar\ (\citealt{Keller14}; unfilled red point), is also the most carbon-rich with an iron abundance upper limit of [Fe/H]~$<-7.1$ and a carbon-to-iron ratio of at least [C/Fe]~$>4.5$. The most metal-poor star currently known,\ \CaffauStar\ (\citealt{Caffau11}; unfilled black point), is the most iron-poor of the carbon-normal stars with [Fe/H]~$= -5$, an upper-limit [C/Fe]~$<0.7$, and metallicity $Z = 10^{-4.35}\,\Zsun$. The shaded region with carbon abundance thickness $-3.8<$~[C/H]~$<-3.3$ represents the abundance space locus where the stars formed from our simulated gas clump would be found. The extent of the carbon-iron ratio in the shaded region, $1 <$~[C/Fe]~$<4$, models uncertainty in both the unknown supernova yield and the potential hydrodynamical enrichment bias that could make the carbon-to-iron ratio in the captured ejecta differ from that of the bulk supernova ejecta \citep[see][]{Sluder16}. The large scatter among the stars in this part of abundance space reflects these theoretical uncertainties and implies a significant degree of stochasticity in the earliest enrichment \citep{Webster16}.

\section{Summary}
\label{sec:summary}

We conclude by briefly reviewing our main results.  This work demonstrates that extremely metal poor ($Z\sim 10^{-4}$--$10^{-3}\,Z_\odot$) stars can form in cosmic minihalos enriched by single, low-energy core collapse supernovae from Pop III star progenitors.  Incomplete photoionization of the Pop III star's host halo and marginally effective carbon and oxygen fine structure line cooling allow extremely metal poor stars to form only $\sim10\,\text{Myr}$ after the Pop III supernova explosion.  

The time scale for recycling ejecta from the first supernovae into second-generation stars depends critically on the presence of neutral gas clouds that can entrain ejecta from the passing blast wave.  The compression and deposition of metals onto the cloud enhances cooling and ultimately leads to gravitational collapse to form new metal-poor stars on shorter timescales, and at higher metallicities, than through the fallback of enriched ISM and IGM returning into the dark matter halo center.  While this simulation did not have sufficient resolution to properly track a time-dependent ionization front eating into a photoevaporating cloud, analytical considerations give us reason to believe that gas with properties similar to that of the birth cloud can survive inside the \hii region long enough to outlive a Pop III star \citep{Bertoldi89,Bertoldi90}.

The survival of a neutral gas cloud in an \hii region is characterized by $\delta$, the initial width of the ionized cloud layer relative to the cloud radius, which scales with the ratio of the photoionization parameter $\xi$ to the neutral hydrogen column density $N_\mathrm{H} \sim r_\mathrm{cloud} n_\mathrm{H}$ \citep{Bertoldi89}.  In terms of the initial cloud properties and source flux, this dimensionless factor becomes
\beq
\delta = 5.23 \times 10^{-2} \left( \frac{S_{49}}{n_3^2 r_{\pc} R_{\pc}^2} \right) ,
\label{eq:delta}
\eeq
where $S_{49} = Q / 10^{49}$ photons per second, $n_3 = n_\mathrm{H} / 10^3 \cmcubed$, $r_{\pc}$ is the cloud radius in parsecs, and $R_{\pc}$ is the distance from the cloud to the ionizing source in parsecs \citep[see Eq. 2.8 of][]{Bertoldi89}.  Clouds with $\delta > 1$ are quickly vaporized, whereas clouds with $\delta < 1$ are evaporated more slowly and survive for some time.  For reference cloud density $10^3\,\cmcubed$ and radius $1\,\pc$ as well as initial distance $1\,\pc$, $\delta = 0.25$ and we expect the cloud to evaporate slowly (in our case $S_{49} = 4.795$). For a cloud that remains at a fixed distance from the star, the evaporation time can be derived from Eqs. 3.27 and 4.10a of \citet{Bertoldi90}
\beq
t_{\mathrm{evap}} \approx 7.43 \times 10^7 \left( \frac{R_{\pc}^2 n_3^2 r_{\pc}^6}{S_{49} T^3} \right)^{1/5} \mathrm{years} .
\label{eq:t_evap}
\eeq
Given the initial cloud parameters above and temperature $T = 200\,\mathrm{K}$, the evaporation time $t_{\mathrm{evap}} = 2.26\,\Myr$ is only about $2/3$ of the lifetime of a $60\,\msun$ Pop III star. However these authors note, and we also observe in our simulation, that the ionized gas pressure force can push the cloud out by several tens of parsecs. Displacement of the cloud from the star decreases the incident flux with the square of the increasing distance.  If the cloud is pushed away to only $R_{\pc} = 10\,\pc$ it would have an evaporation time $t_{\mathrm{evap}} = 5.68\,\Myr$, several million years longer than the host star's lifetime.  This model suggests that surviving neutral clouds are generic in \hii regions of single $\lesssim 60\,\msun$ Pop III stars in minihalos.  If present, these clouds can become the birth sites for the first metal-poor stars.

The enrichment mechanism we identify here differs from the well explored scenario involving ultra-energetic (e.g., pair instability) Pop III supernovae \citep{Wise08,Greif10}.  Since pair instability supernovae (PISNe) synthesize large ($\sim 50\,M_\odot$) metal masses and inject them with large ($\sim 10\text{--}100\times10^{51}\,\text{erg}$) initial kinetic energies, the formation of very metal poor PISN-enriched stars would require metal dilution, through hypothesized pervasive hydrodynamic mixing, over a large cosmic volume.  Even then, the abundances of PISN-enriched stars would differ from those in the extremely metal poor stars observed to date \citep[e.g.,][]{Frebel13,Yong13}.\footnote{Tentative PISN enrichment pattern evidence has been found only in one less-than-extremely metal poor ($[\text{Fe}/\text{H}]\approx-2.5$) star \citep{Aoki14}.}

In contrast with the massive progenitor, ultra-energetic explosion Pop III enrichment scenario, we find that the first metal-enriched gravitationally collapsing pre-stellar clump acquires its extremely low metallicity through superficial deposition by a transient supernova blast wave.  The blast wave deposits a small mass of metal-rich supernova ejecta on the surface of the clump's progenitor cloud.  Correcting for numerical artifacts, we expect that chemical homogenization in the pre-stellar clump does not occur before the clump has entered runaway gravitational collapse when turbulence, if present, can facilitate compositional mixing.  

We find that cooling by carbon and oxygen fine structure line emission is an effective accelerator of the formation and collapse of the metal-enriched pre-stellar clump.  We also find that the clump enters collapse on an elliptical orbit around the dark matter halo gravitational center.  This implies that extremely metal poor stars can be born with a significant initial radial velocity dispersion.  If the halo survives as a dwarf galaxy, stellar dynamical evolution could isotropize the velocity dispersion tensor on Gyr time scales. The radial orbital extent may be preserved and would be reflected in the half-light radii of, e.g., the UFD satellite galaxies, local systems within observational reach of stellar archaeological studies.

The confluence of ongoing large stellar archaeological surveys such as
Gaia and SkyMapper with high-resolution cosmological simulations
of chemically primitive star formation environments is opening up
a powerful window into the nature of the first stars and galaxies.
In particular, we can now begin to understand the multiplicity of
the initial chemical enrichment process, single vs. multiple supernova
pathways, and formulate robust empirical near-field cosmological tests
of our emerging theoretical framework.

\section*{Acknowledgments}

The \textsc{flash} code was in part developed by the DOE-supported Flash Center for Computational Science at the University of Chicago. Many of the visualizations were made with the \textsc{yt} package \citep{Turk11}. M.M.'s work was performed in part at the Aspen Center for Physics, which is supported by National Science Foundation grant PHY-1066293. The authors acknowledge the Texas Advanced Computing Center at The University of Texas at Austin for providing HPC resources under XSEDE allocation TG-AST120024. This study was supported by the NSF grant AST-1413501.

\bibliographystyle{mn2e}
\bibliography{metalpoor}

\label{lastpage}
\end{document}